\documentclass[aps,preprintnumbers,nofootinbib,superscriptaddress]{revtex4}
\makeatletter
\renewcommand\@biblabel[1]{#1.}
\makeatother

\usepackage{graphicx}
\DeclareGraphicsExtensions{.pdf}

\usepackage{amsfonts,amssymb,amsmath}
\usepackage{amsthm}

\newcommand{\comment}[1]{}
\newcommand{\ket}[1]{\left |  #1 \right\rangle}
\newcommand{\bra}[1]{\left \langle #1  \right |}

\newcommand{\braopket}[3]{\langle #1 | \, #2 \, | #3 \rangle}

\theoremstyle{plain}

\theoremstyle{definition}

\begin{document}

\title{Secure and Robust Transmission and Verification of Unknown Quantum States in Minkowski Space}

\author{Adrian \surname{Kent}${}^{*}$}
\affiliation{Centre for Quantum Information and Foundations, DAMTP, Centre for
  Mathematical Sciences, University of Cambridge, Wilberforce Road,
  Cambridge, CB3 0WA, U.K.}
\affiliation{Perimeter Institute for Theoretical Physics, 31 Caroline Street North, Waterloo, ON N2L 2Y5, Canada.}

\author{Serge \surname{Massar}}
\affiliation{Laboratoire d'Information Quantique, CP 225, Universit\'{e}
Libre de Bruxelles, AV. F. D. Roosevelt 50, 1050 Bruxelles, Belgium}
\author{Jonathan \surname{Silman}}
\affiliation{Laboratoire d'Information Quantique, CP 225, Universit\'{e}
Libre de Bruxelles, AV. F. D. Roosevelt 50, 1050 Bruxelles, Belgium}

\begin{abstract}

  An important class of cryptographic applications of relativistic
  quantum information work as follows.  $B$ generates a random qudit
  and supplies it to $A$ at point $P$.  $A$ is supposed to transmit it
  at near light speed $c$ to to one of a number of possible pairwise
  spacelike separated points $Q_1 , \ldots , Q_n$.  $A$'s transmission
  is supposed to be secure, in the sense that $B$ cannot tell in
  advance which $Q_j$ will be chosen.  This poses significant
  practical challenges, since secure reliable long-range transmission
  of quantum data at speeds near to $c$ is presently not easy. Here we
  propose different techniques to overcome these diffculties.  We
  introduce protocols that allow secure long-range implementations
  even when both parties control only widely separated laboratories of
  small size. In particular we introduce a protocol in which $A$ needs
  send the qudit only over a short distance, and securely transmits
  classical information (for instance using a one time pad) over the
  remaining distance.  We further show that by using parallel
  implementations of the protocols security can be maintained in the
  presence of moderate amounts of losses and errors.

\end{abstract}

\maketitle
${}^{*}$  Correspondence to a.p.a.kent@damtp.cam.ac.uk
\newpage
%\section{Introduction}
\label{sec:intro}

Quantum theory and the relativistic no-signalling principle 
both give ways of controlling information, in the sense
that someone who creates information somewhere in
space-time can rely on strict limits both on how much information
another party can extract and on where they can obtain it. 
While standard quantum cryptography
(e.g.
\cite{wiesner,BBeightyfour,ekert,qsecrets,anonqcomm,qmultiparty}) 
uses only the properties
of quantum information, an increasingly long list of applications illustrate the
added cryptographic power of the relativistic no-signalling principle, either alone
(e.g. \cite{power,kentrel,kentrelfinite,colbeckkent}), or when combined with quantum information
(e.g. \cite{taggingpatent,malaney,kms,chandranetal,buhrmanetal,kenttaggingcrypto,beigikoenig,BHK,BKP,AGM,
AMP,ABGMPS,PABGMS,McKague,MRC,Masanes,HRW2,MPA,HR,ColbeckThesis,PAMBMMOHLMM,ckrandom,DIBC,
nosummoning,bcsummoning,otsummoning,bcmeasurement}).  

A new relativistic quantum cryptographic technique was 
recently introduced, inspired by the no-summoning theorem
\cite{nosummoning}, in which one agency (Alice or $A$) sends a quantum state,
supplied by and known to
another agency (Bob or $B$) but unknown to $A$, at light speed $c$ in one of 
a number of possible directions.  We use the term agency here to 
stress that Alice and Bob are not single isolated individuals: they
have representatives (who we assume are all loyal and
act according to the instructions of their agency) 
distributed at various points in space-time.
The task is securely implemented if the chosen direction is concealed
from $B$ until $A$ chooses to return the state.   
This gives, inter alia, a provably unconditionally secure protocol for 
bit commitment
\cite{bcsummoning} and a way of transferring data at a location unknown
to the transferrer \cite{otsummoning}.

Another class of applications arises in scenarios where the unknown
state is a quantum authentication token supplied either by $B$ or by a
third party $C$.   $A$ may wish
to use this token, by transferring it to $B$, at a space-time 
location $Q_i$ that she chooses, without giving $B$ advance 
notice of her choice.  For example, the token here might be 
a quantum password authenticating access to a site, or quantum
money used to execute a trade.   

In their simplest forms, all these applications 
require $A$ to propagate the quantum state securely from one point $P$
to one of a number of points $Q_1 , \ldots , Q_n$ in its causal future,
in such a way that $B$ cannot obtain any advance information about $A$'s choice of
$Q_i$.   We first discuss ways of implementing
this task of secure unknown state transmission to a
hidden location.   

Next we consider an extended task where $A$ also wants to allow 
$B$ to verify that the state was sent to the chosen $Q_i$, at or 
after the point $Q_i$.  If the state was indeed securely transmitted,
this can in principle be achieved by returning it to $B$ at $Q_i$, where $B$ can measure
the returned state and test whether it is the state he originally
provided.  
As we will see, if the state originally supplied is a pure qudit known to
$B$, Alice's probability of passing this test at 
two distinct specified sites 
is $O( d^{-1} )$, where $d$ is the dimension of the Hilbert space. 
We can thus ensure both that $A$'s transmission is secure against $B$ 
until the state is returned and that $B$ can be near-certain of
which point $Q_i$ was chosen after the state is returned.  

However, secure (or even insecure) quantum transmission from $P$ to
the $Q_i$ may not always be possible for $A$.   Even in a future
world in which reliable long-range quantum transmission technology
exists, one can easily imagine scenarios in which $A$ (a user of the 
infrastructure who has relatively few 
resources) does not have such technology, 
while $B$ (an infrastructure provider with large resources) does.     
We describe here strategies for such scenarios, in which $A$ and $B$ 
exchange classical
and quantum information  so as to have the same cryptographic 
effect -- i.e., $A$ 
chooses one of the $Q_i$, a choice that $B$ can verify
at or after the relevant $Q_i$ but not earlier -- without 
$A$ necessarily actually transmitting any unknown states
to any of the $Q_i$.  
We also describe a variation which achieves this result even
when neither party has reliable long-range quantum transmission.
 
Furthermore, real life implementations must also allow for losses and
errors.  Here we show that relativistic quantum cryptography can be
made secure even in the presence of moderate levels of losses and/or
errors.

The results presented here show that there is considerable flexibility in
the implementation of relativistic quantum cryptographic protocols.
This makes the prospect of experimental implementation of such
protocols much more realistic, thereby bringing them closer to
practical applications.

We suppose throughout that $B$ does in fact carry out the optimal
verification measurements on states returned to him.   The reason
for assuming this is to define a quantifiable measure of security
against $A$.  In practice, $B$ may actually verify only some, or none, of the returned
states in any given run, depending on the ultimate application.

\section*{Results}
\subsection*{Secure unknown state transmission to a hidden location}

Theoretically, the standard assumptions for mistrustful
cryptography give a way to justify assuming that $A$ can securely
transmit an unknown state from $P$ to any $Q_i$ of her choice in the
causal future of $P$, without $B$ learning the choice of $Q_i$ in advance.
The reason is that we must, in any case, assume that $A$ has a ``secure
laboratory'' -- some region shielded from $B$ in which $A$ can carry out any classical
or quantum operation securely.  If we take $A$'s laboratory to include
linear channels between $P$ and $Q_j$ for each $j$, her quantum
transmissions along these channels are, by definition, secure.

In practice, however, it may not always be so easy to justify this
security assumption. 
Keeping a long quantum channel physically secure against
all eavesdropping is problematic.  Indeed, even insecure high fidelity
terrestrial quantum transmission over more than $100$km or so is currently difficult.   
A further practical problem is that one would ideally like no
constraint on the $Q_i$ except that they lie in the causal future of
$P$: in particular one would like to be able to implement the task 
when the $Q_i$ are (nearly) lightlike separated from $P$. 
However, the presently most advanced quantum
communication technology involves sending light pulses along standard
fibre optic cables.  These have relatively high refractive index, so
that the transmission speeds are typically only about $\frac{2}{3} c$. 
Sending light pulses through air or free space is an alternative, and
allows transmissions at speeds close to $c$, but keeping such
transmissions secure over long ranges may be harder
still.

(Perhaps photonic crystal fibres may in the future offer another
way of sending light pulses at speeds close to $c$: even if so, the
problem of secure long range transmission remains.) 

Here we propose ways around these problems, in the form of
implementations that do not require long physically
secure quantum channels, or in some cases even long range quantum communication.   

To simplify the discussion,
we start with the idealized assumptions that all local 
classical and quantum operations 
in the protocol can be implemented perfectly and instantaneously
and that all classical and quantum signals can be sent at in vacuo 
light speed. 
We discuss errors, losses and delays later.
We also start with the assumption that both parties can verify
that any qudit involved in the protocol really is a qudit, i.e. that
it is completely characterised as a quantum state in a prescribed
$d$ dimensional Hilbert space and carries no hidden information in other
degrees of freedom.  While this assumption is quite standard in 
the quantum cryptographic literature, it is obviously questionable.
We discuss later how it can be guaranteed. 

\subsubsection*{Secure quantum channels}

We are interested in three types of secure quantum channel:
\\

{\bf S1. Physically secure quantum transmission channels} \qquad  As
already noted, it is standard (and necessary: without such
  an assumption, cryptography is effectively pointless), in considering
  cryptography involving mistrustful parties, to assume that each
  party has at least one secure ``laboratory'' -- a region of space (or, more
generally, space-time) within which they can carry out any operations
they wish with an assumed guarantee of privacy.    In particular, the
parties can transmit quantum states securely within their laboratory. 
One theoretically legitimate way of ensuring that a party has a secure
quantum channel between space-time points $P$ and $Q$ is thus simply to
assume that their laboratory contains a timelike or lightlike path
between the points.
\\
{\bf S2. Quantum teleportation between secure sites} \qquad If $A$ has predistributed a maximally entangled $d$-dimensional state 
$\sum_{i=1}^d \ket{i}_P \ket{i}_Q$ between secure sites $P$ and $Q$, she
can teleport an unknown qudit $\ket{\psi}_P$ from $P$ to $Q$ by carrying out
a generalized Bell measurement on the two qudits at $P$, transmitting the 
outcome $i$ (which ranges from $1$ to $d^2$) classically over a public
channel to $Q$, and applying
an appropriate unitary $U_i$ to the qudit at $Q$.  
The outcome $i$ carries no information about $\ket{\psi}_P$: all outcomes
are equiprobable.    Thus, $\ket{\psi}_P$ is securely transmitted to $Q$,
even though the classical channel is public.  

Note also that if $A$
has predistributed maximally entangled $d$-dimensional states 
between $P$ and $Q_j$, for a range of $j$, she can choose which
$Q_j$ to teleport the qudit to.   If she broadcasts the classical
teleportation data, not only is $\ket{\psi}_P$ securely transmitted to $Q_j$,
but the identity of $Q_j$ is kept secret from $B$, unless and until 
$A$ chooses to reveal it, for instance by returning $\ket{\psi}$ to
$B$ there.  
\\
{\bf S3. Transmitting randomised quantum states} \qquad Another way of
securely transmitting 
$\ket{\psi}_P$ is first to apply a randomly chosen teleportation
unitary at $P$, creating the state $U_i \ket{\psi}_P$, then to send
this state over a (not necessarily secure) quantum channel, 
while separately sending the classical data $i$ over a secure
classical channel along (essentially) the same path.   
The state $\ket{\psi}$ is then regenerated by applying $U_i^{\dagger}$
at the end of the path.   The quantum transmission
carries no information about $\ket{\psi}$, since 
\begin{equation}
\sum_i U_i \ket{\psi} \bra{\psi} U_i^{\dagger} = \frac{1}{d} I_d \, . 
\end{equation}
By assumption, the classical data $i$ are sent securely, and so also
give $B$ no information about $\ket{\psi}$.   

Hence, if $A$ wants to send $\ket{\psi}$ from $P$ to a site $Q_j$, and keep
the identity of $Q_j$ secret, she need simply generate appropriate dummy
transmissions from $P$ to $Q_k$, for each $k \neq j$.   One way of
doing this is to send randomly chosen qudits from $P$ to each $Q_k$
($k \neq j$) while securely broadcasting the classical data $i$ from $P$ to all 
the sites $Q_k$ (including $Q_j$). 

In each case, $A$ uses a secure classical channel to tell
her agents at each $Q_k$ whether they have received the real state
or a dummy.   The secure classical channels we consider here and below
could, for instance, be created by previously distributed secret keys 
that allow encryption of classical information followed by public
broadcasting.  

In theory, any of the secure channels $1-3$ could be used for long
distance relativistic quantum cryptographic protocols.  
Presumably technology will ultimately make long range quantum channels of
types $2$ and $3$
practical and reliable. 
However, at present, we do not have reliable practical teleportation,
nor reliable long-term quantum state storage for teleportation with predistributed
entanglement, and the necessary quantum transmissions for type $3$
channels become increasingly challenging as the channel length (in the lab rest frame) 
increases.   Ensuring that a quantum transmission channel of type
$1$ is physically secure is also challenging: it requires resources that
scale at least linearly with length, and may simply not be practical or possible over long distances in many
scenarios.  For example, a party may be able to set up secure labs at
well separated locations on Earth and/or satellites, but have no way
of keeping secure any nearly linear quantum channel path between them:
all such paths may go through regions controlled by or open to others.  

\subsection*{Extending a secure quantum channel classically}

We propose the following technique to get around these obstacles.
Suppose that Alice controls some (perhaps small) region around the
space-time point $P$ from which she wishes to send an unknown qudit to one of the 
space-time points $Q_j$ ($j = 1 , \ldots , N$) at (or near) speed
$c$.  Define the space-time points $P'_j$ such that $P$, $P'_j$ and $Q_j$ are
collinear, and $P'_j$ is on the boundary of the region she controls
around  $P$.   In practice, we imagine $P'_j$ will generally be much
closer to $P$ than to $Q_j$.  
Instead of sending a quantum state $\ket{\psi}$ securely from $P$ to $Q_j$,  and returning it to Bob there, Alice 
sends $\ket{\psi}$ securely from $P$ to $P'_j$, applies a random
teleportation unitary $U_i$ somewhere en route, and returns the state $U_i
\ket{\psi}$ to Bob at $P'_j$.   She transmits the classical data $i$ securely
from $P'_j$ (or wherever it was generated) to her lab at $Q_j$, and gives it to Bob (only) there.    
(See Figure 1.) 

\begin{figure}[t]
\centering
\includegraphics[scale=0.5]{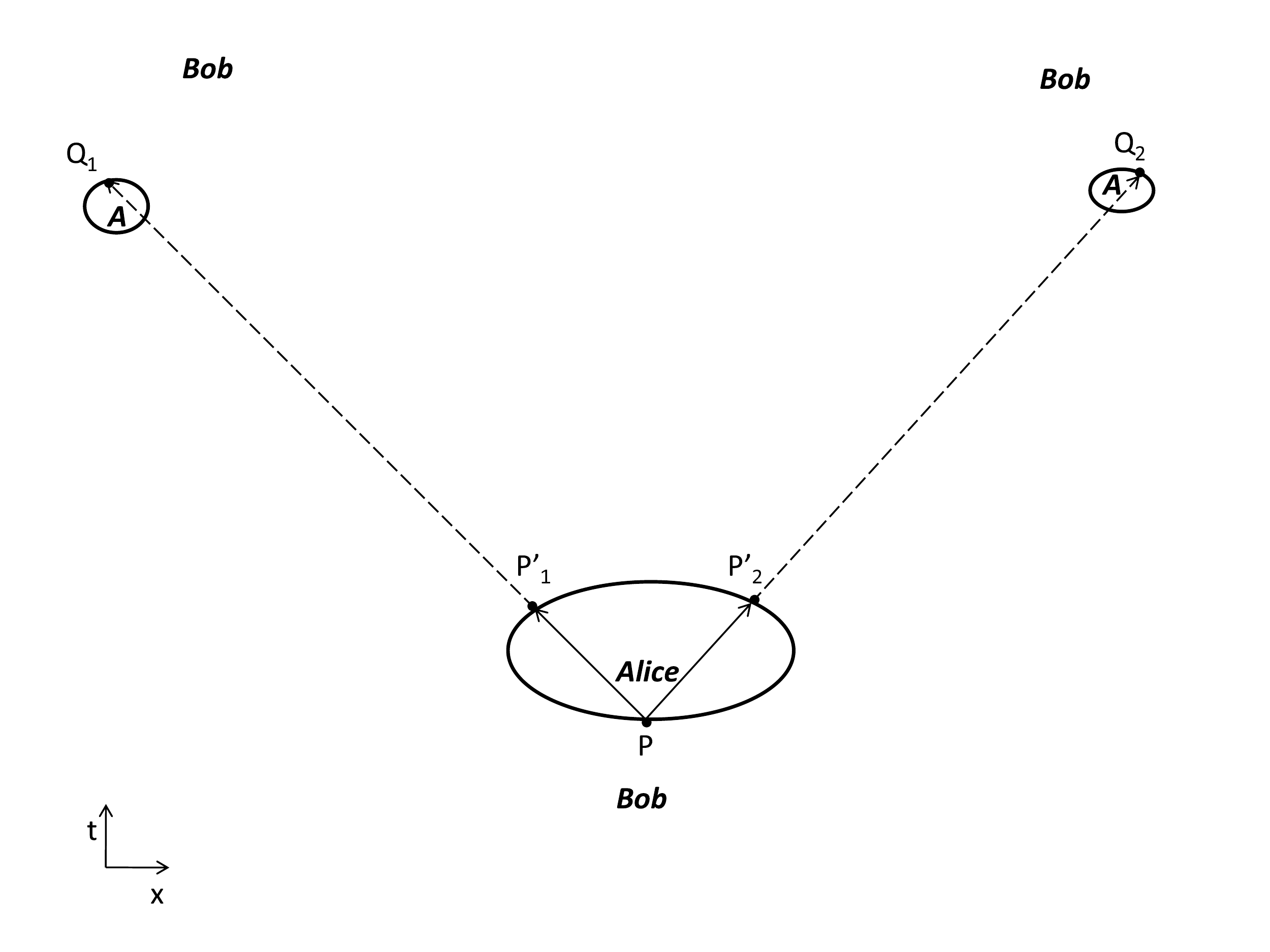}
\caption{Illustration, in $1+1$ dimensions (not to scale), of how quantum channels can be securely extended classically.
Alice controls a laboratory
including on its border the points $P$, $P'_1$ and $P'_2$, and
disjoint laboratories including on their borders the points $Q_1$ 
and $Q_2$ respectively, where $P P'_1 Q_1$ and $P P'_2 Q_2$ are
lightlike lines.   Bob may control the rest of space-time.
Alice receives the unknown state $\ket{\psi}$ from Bob at $P$, 
and transmits it securely to $P'_i$ (where she chooses $i =1$ or $2$).
There
she randomizes it and returns the randomized state to Bob.  The
classical data describing how the state was randomized are 
transmitted along a secure classical channel and returned to Bob at $Q_i$.
Alice 
transmits dummy quantum and classical information and returns 
them to Bob at the corresponding points on the opposite
wing.} 
\end{figure}

Precisely how and where the unitary $U_i $ is applied depends on the type of quantum channel Alice
uses between $P$ and $P'_j$.   In the cases of quantum teleportation, the teleportation measurement generates
$i$ (which Alice sends securely from $P$ rather than $P'_j$ to $Q_j$), and the state $U_i \ket{\psi}$ is created
at $P'_j$ by the teleportation.   In the case of randomised quantum state transmission, Alice generates
the randomised state $U_i \ket{\psi}$ at $P$, transmits it to $P'_j$ and returns it there to Bob, and again
sends $i$ securely from $P$ to $Q_j$.   If the channel from $P$ to $P'_j$ is physically secure, Alice need
not generate $i$ and randomise $\ket{\psi}$ until $P'_j$, although again she can do both at $P$ if she wishes.

In each case, to ensure that Bob does not learn the choice of $Q_j$
prematurely, Alice generates random dummy qudits that are
returned to Bob at $P'_k$ (for all $k \neq j$).  She also either sends random dummy
classical transmissions towards each $Q_k$ from $P'_k$ (for $k \neq j$)
or, if her classical data were already generated at $P$, broadcasts the
classical data securely to all the $Q_k$.  Finally, via secure
classical channels, she tells each of her agents at $Q_k$ whether 
``their'' state is real or a dummy.  

\subsubsection*{Security of classically extended quantum state transmission}

In their original form, the various relativistic cryptographic schemes we consider are secure against 
Alice essentially because the no-cloning \cite{wz,dieks} and no-summoning
\cite{nosummoning} theorems imply that she cannot generate copies of the unknown
state $\ket{\psi}$ at more than one point $Q_j$, provided the $Q_j$
are spacelike separated.   
Moreover, if $P$ is lightlike separated from each $Q_j$, then her actions at $P$ 
essentially determine at which of the $Q_j$ she will be able to
generate $\ket{\psi}$. 
More generally \cite{bcsummoning}, they determine her optimal probabilities $p_j$ of producing states that give Bob a positive
outcome for a projective measurement onto $\ket{\psi} \bra{\psi}$ at
$Q_j$, 
and for a fixed number of sites these 
satisfy $\sum_j p_j \leq 1 + O ( d^{-1} )$.    

Perhaps counterintuitively, the same security argument also holds for
the classically extended versions of the scheme described here.  The
reasoning is simple.  When Bob receives the state $U_i \ket{\psi}$ at
$P'_j$, he {\it could} transmit it at light speed from $P'_j$ to $Q_j$
and, upon receiving the classical information provided by Alice at
$Q_j$, apply the inverse unitary $U^{\dagger}_i$. This procedure
produces at each site $Q_j$ a quantum state. Compared to the situation
where Alice has physically secure quantum transmission channels from
$P$ to the $Q_j$ (case {\bf S1}), this procedure is more restricted
for Alice, and hence offers her less cheating strategies. Therefore
the bounds that apply in case {\bf S1} also apply for the present
protocol, and in particular the optimal probabilities of these states
passing Bob's tests must satisfy $\sum_j p_j \leq 1 + O ( d^{-1} )$.

These classically extended schemes are also clearly secure against Bob, since he possesses no information 
correlated with $\ket{\psi}$ until he can combine both $U_i \ket{\psi}$ and $i$, i.e. until the point $Q_j$ 
where the state is returned.   

\subsubsection*{Other schemes and resource tradeoffs}

In these classically extended schemes, Alice needs to be able to receive an unknown state, randomize it and
transmit it a short distance securely at light speed along her chosen path, return it to Bob, and transmit 
the classical randomization data at light speed securely over a long distance.    Short range secure
transmission of a quantum state is, clearly, easier than long range
transmission.
  Reliable long range secure transmission of
classical data can be easily achieved, using pre-shared one-time pads
(or, if the range is within the limits of current technology, quantum key distribution).   
These schemes thus appear practically advantageous for Alice.   

Note, however, 
that for many cryptographic applications
Bob must be able to verify that the points $P'_j$ at which the
state is returned are space-like separated.
For example, this is necessary in the bit commitment schemes of 
Ref. \cite{bcsummoning}, which require Alice to be forced to choose
$Q_j$ (and hence $P'_j$) at or before the point $P$.  (To be precise,
Alice is required to choose a probability distribution over $j$, 
since she can commit to a superposition of bits.)  
Any other cryptographic application that requires
Alice to be prevented from deciding the value of $j$ at a point
in the future of $P$ similarly requires the $P'_j$ to be space-like
separated: if they are not, she can sequentially send the quantum
state between non-space-like separated $P'_j$ and decide en route
at which site she will return the state. 
The need for space-like separated $P'_j$ poses its own practical
challenges, requiring more precise timings, and shorter transfer
delays, as the separations grow shorter.
   
Note also that, instead of receiving the quantum state $\ket{\psi}$ at $Q_j$
and measuring it there, Bob now receives separately $U_i \ket{\psi}$ at $P'_j$ and $i$ at $Q_j$.  
He has more than one option, depending on the available technology and on how quickly (where in spacetime)
he needs to verify the unveiled commitment.   For example:

{\bf B1.} \qquad Bob can transmit $U_i \ket{\psi}$ from $P'_j$ to $Q_j$, apply $U^{\dagger}_i$ at $Q_j$, 
and check there that the resulting state is indeed $\ket{\psi}$.
This requires Bob rather than Alice to have a long range
quantum channel, an advantage in the scenarios we envisage in which
Bob's technology is better.  A further advantage is that Bob's long range channel need
not necessarily be secure.   Alice would gain no cheating advantage, compared to the original version of the
  scheme with long-range quantum transmissions, by being able to
  tamper with this state en route -- in the original scheme, she
  controls the state throughout the path from $P$ to $Q_j$. 

{\bf B2.} \qquad Bob can store $U_i \ket{\psi}$ at the space coordinate of $P'_j$, transmit $i$ there from
$Q_j$, apply $U^{\dagger}_i$ there, and check that the resulting state is $\ket{\psi}$.
This requires Bob to have quantum state storage.   It also delays his
verification of the state, which is a disadvantage in some contexts.
For example, if the state is used as an authentication token, Bob
cannot immediately verify Alice's authentication at $Q_j$.   
However, it would be advantageous in a possible scenario in
which reliable local quantum state storage turns out to be easier than 
reliable long range quantum transmission. 

These two options have one very clear advantage in the near-ideal
case where Alice and Bob's errors and losses can be made  
small through error correction or advanced technology. 
In this case the quantum information encoded in the unknown state
remains essentially intact.  Bob can thus make use of the token
as he wishes.  For example, he could return the token to Alice 
if, at the relevant point $Q_j$, she decides she wishes
not to use it, but to propagate it to another site in the causal
future of $Q_j$.  Alternatively, 
he could pass it on to a third party for storage
or testing.

\subsection*{Errors, losses and delays}\label{errorsetal}

In realistic implementations of protocols using any of the secure
channels $S1-S3$, $A$ and $B$'s channels and devices
will introduce some errors, their channels will suffer losses, and
Bob's detectors will also cause losses as well as detection errors.
We now show that, in principle, provided the levels of losses and
errors are not too great, they can be securely countered by
redundant implementations of the protocol.  

The idea here is that, instead of providing $A$ with one random qudit at $P$, $B$ provides her with $N$ 
independently chosen random qudits in short time sequence (so they
all arrive close in time to the spacetime point $P$). 
$A$ is supposed to send all of them to (points correspondingly close
in space-time to) her chosen site $Q_j$ and return them to $B$ there.
$B$ accepts the site as chosen provided that he receives and 
verifies a sufficient fraction of the qudits. 

This protocol is more robust than the original protocol because Bob
will not abort even if a small fraction of the particles are lost or
corrupted by noise. We show below that this protocol can tolerate
losses of up to $(\frac{1}{2} - \frac{1}{d+1})$ of the particles.  For definiteness we consider
only the case where $A$ tries to pass Bob's statistical tests at two
distinct sites, $Q_1$ and $Q_2$.  The argument generalizes simply to
the multi-site case using the fidelity bound
\cite{gm,bem,keylwerner,werner} on symmetric $1 \rightarrow n$ approximate
cloning.   

Recall first the situation when Alice is provided with a single qudit. She returns to Bob at sites $Q_1$ and $Q_2$ the states $\rho_{1}$ and $\rho_{2}$ respectively. 
The probabilities  of passing Bob's test  at the two sites are
\begin{equation}
p_{1} = \braopket{\psi}{\rho_{1}}{\psi} \, , \qquad 
p_{2} = \braopket{\psi}{\rho_{2}}{\psi} \, . 
\end{equation}
They 
obey \cite{bcsummoning}
\begin{equation}\label{bound} p_{1} + p_{2}  \leq  1 + \frac{2}{d+1} \, . \end{equation}

Now suppose Alice is given the $N$ independent random qudits
$\ket{\psi_1} , \ldots, \ket{ \psi_N}$, and carries out a collective operation
producing outputs $ \rho_{i1}$ and $\rho_{i2}$ for $\ket{\psi_i}$,
which are returned to Bob for testing at the respective sites. 
A complication in analysing this situation is that Alice can act
collectively on all $N$ qudits, which 
in principle might increase her cheating probability compared to 
strategies involving only individual qudit operations.  We now
show this is not the case.  

We suppose that Bob  tests all $N$ pairs of returned qudits 
in succession at sites $Q_1$ and $Q_2$. 
Denote by $N_1\leq N$ and $N_2\leq N$  the number of qudits that pass Bob's test at sites $Q_1$ and $Q_2$ respectively. Bob accepts at site $Q_i$ if $N_i>\frac{N}{2}( 1 + \frac{1}{d+1}+\epsilon )$.
Let $P_{i}^{N}=\mathrm{Prob}\left(N_i>N ( \frac{1}{2} + \frac{1}{d+1}+\epsilon )\right)$ be the probability that Bob accepts at site $Q_i$.
We show that the sum of the probabilities of acceptance at both sites is bounded by 
\begin{equation}
P_{1}^{N}+P_{2}^{N}\leq  1+\exp\left[-\frac{N\epsilon^{2}}{2(1+\frac{2}{d+1})^2}\right]
\label{secureN}
\end{equation}
which is the analog of eq. (\ref{bound}) and establishes security against Alice. 

We note that security against Bob is as before, since he does not have access to any information until Alice provides him with the reveal information. 

The proof of eq. (\ref{secureN}) is carried out in two steps. The
first step essentially shows that collective operations do
not help Alice. We prove the following. Suppose that 
Bob's tests on $\rho_{i1}$
and 
$\rho_{i2}$ for $ 1 \leq i \leq k-1$ produce some results. Then, conditional on these results, 
the fidelities 
$p_{k1}$ and 
$ p_{k2} $ on the $k$-th pair of returned qudits
 must satisfy 
\begin{equation} \label{Nbound} 
p_{k1} + p_{k2} \leq 1 + \frac{2}{d+1} \, . \end{equation}

To see this, suppose that there were some collective operation $O$
for which, conditioned on some set $D$ of results for $1 \leq i \leq k-1$,
Alice could violate this bound.  
She could then proceed as follows:
(i) create a random set of qudits $\ket{\psi_i}$ for $1 \leq i \leq k-1$
together with a maximally entangled state of two qudits;
(ii) apply the operation $O$ on the $\ket{\psi_i}$ together with one
of the entangled qudits; 
(iii) test the first $(k-1)$ outputs; 
(iiia) if she does not obtain distribution $D$ of the results, return to (i)
generating a new set of qudits. 
(iiib) if she obtains distribution $D$, apply a teleportation
protocol using $\ket{\psi_k}$ and the other entangled qudit, use the
teleportation measurement result to define a teleportation unitary $U$
in the standard way, then apply  $U$ to the two outputs corresponding
to $\ket{\psi_N}$, and return these rotated outputs to Bob for testing.

The probabilities of Bob's test
outcomes after step (iiib) are independent on whether the teleportation was carried
out before or after the cloning operation, and so must violate
(\ref{Nbound}).  
But this is a contradiction, since Alice now has a generally
applicable protocol for violating the universal bound
(\ref{Nbound}) for cloning a single unknown state. 

The second step uses large deviation results for martingales, and in particular the Azuma-Hoeffding
inequality \cite{gs,azuma,hoeffding} to prove eq. (\ref{secureN}). 
This more technical step is given at the end of the paper.

Note that security for redundant protocols always requires a total
error and loss rate, for Alice's operations, of less 
than $\frac{1}{2}$.   If the allowed total rate is greater than $1/2$, 
a technologically advanced Alice whose devices have no losses or
errors could cheat by sending $1/2$ the states to each of $Q_1$ and
$Q_2$ and later choosing the site at which she will return the
correct states to Bob.  
This choice may depend on information that may have become
available at both $Q_1$ and $Q_2$, allowing her actions at the
two sites to be coordinated.  For example, it could depend
on some event that will happen either at $Q_1$ or $Q_2$ but not
both and that Alice cannot predict in advance. (See
Ref. \cite{qtasks} for a formal general discussion of security
based on an oracle input model.   An independent discussion of
another proposed security model can be found in Ref. \cite{kthw}.)

On the other hand, the above argument shows that for sufficiently large $N$ a loss
rate lower than $\frac{1}{2} - \frac{1}{d+1}$ suffices.   
Even the weaker of these bounds -- total error and loss rate less than
$\frac{1}{2}$ -- appears practically challenging with current technology. 

Delays are another practical issue: realistically, $A$ and $B$'s state exchanges and quantum operations
will not be instantaneous, and their channels will transmit
quantum states at speeds slower than $c$.  As noted
elsewhere\cite{nosummoning},
such delays affect the details of $B$'s security guarantee -- he is no
longer guaranteed that $A$ was effectively committed to a choice
$Q_j$ (or probability distribution of choices) precisely at the point 
$P$, for example.  
(As with all technologically unrestricted quantum protocols 
\cite{bccc,classcert}, Alice can create a quantum superposition of
commitments -- in this case at, or in the non-ideal case near, $P$.      
Nonetheless, her optimal probabilities for successful unveiling are
constrained by Eq. \ref{Nbound}.)
However, so long as they are small compared
with the spacelike separations of the $Q_j$, $B$ is still
guaranteed that $A$ was effectively committed to a choice (or
probability distribution of choices) of
site $Q_j$ in advance, at a point relatively close to $P$. 

Note also that, for secure implementations of variations $B1$ and $B2$ to be possible,
Alice needs to receive and return $N$ states in
times short compared to the short separations between $P$ and the
sites $P'_j$.  

\subsection*{Long-range implementations when neither party has long
  distance state transmission or state storage} 

Redundant implementations also allow security even where neither party has reliable long distance
state transmission or state storage.  For example: 

{\bf B3.} \qquad  Suppose again that the unknown state is replaced
by $N$ independent states, generated by Bob.    Alice receives these $N$ labelled independent random
states $\ket{\psi_k}$ in quick succession at (approximately) the same point $P$,
and is required to return the ordered set of states $\{
U_{i_k} \ket{\psi_k} \}$ to Bob at $P'_j$ and the ordered set of numbers $\{
i_k \}$ to Bob at $Q_j$.  Bob can then test her claimed choice of $Q_j$ statistically by
carrying out suitable randomly chosen measurements on the states
$U_{i_k} \ket{\psi_k}$, and comparing the results to the list $\{ i_k
\}$ at any point where he can combine both sets of data: in particular
he can do so at $Q_j$.  

For example, one simple test strategy is for Bob to take $N = M d^2$ for 
$M \gg 1$ and to make random guesses $i'_k$ of the value of each $i_k$, carry
out measurements at $P'_j$ to test whether the returned state is
$U_{i'_k} \ket{\psi_k}$ (for each $k$), and later compare these
results to the values $i_k$, focussing on the $\approx M$ cases where he guessed
correctly (i.e. $i'_k = i_k$).  The probability that Bob guesses
correctly at least (say) $\frac{2M}{3}$ times is $ 1 - \eta (M)$, where $ \eta (M)
\rightarrow 0 $ as $M \rightarrow \infty$.  $M$ is chosen so that
$\frac{2M}{3}$ is suitably large for the security argument above to
apply given the level of errors and losses encountered, while $\eta (M)$ is suitably small.    

This variation still requires a total error and loss rate lower than $1/2$, and
so still appears practically challenging with current technology.
However, as it requires neither quantum state
storage nor long range quantum transmission from either party,
it may be implemented in practice
sooner than $B1$ or $B2$, as it requires only states of small dimension $d$ to be 
created and manipulated, with no more advanced quantum technology needed
by either party. 

Note again, though, that for a secure implementation to be possible
Alice needs to receive and return $N$ states in
times short compared to the short separations between $P$ and the
sites $P'_j$.   Also, the quantum information
encoded in the states is effectively destroyed by Bob's measurements, so this
variation of the protocol cannot generally be combined with applications in
which some quantum states supplied by Bob are subsequently used for
another purpose.

\subsection*{Side channels and related security issues}

As noted above, a quite general potential concern in quantum
cryptography is that an adversary may exploit physical degrees of
freedom other than those prescribed by the protocol.  For example, in
the protocols considered here, Bob is supposed to provide a randomly
chosen qudit, which might for instance be supposed to be encoded in
photon polarizations.  He could, however, use a side channel and
surreptitiously encode extra information in other degrees of freedom,
such as position/momentum, energy/time, or number/phase.  If Alice
then sends the photons along physically insecure channels, she is in
principle vulnerable, even if the polarization degree of freedom is
appropriately randomized.  If Bob can carry out a nondestructive
measurement of the other degree of freedom, he can identify the
channel carrying the photons he supplied, without detection.

A (theoretically) simple way for Alice to counter
this was first noted by Lo and Chau \cite{lochaurobot} in considering an analogous security problem
in quantum key distribution.   Alice can prepare her own maximally
entangled pairs of qudits, in separate secure (but closely adjacent)
labs in the neighbourhood of $P$.    She teleports the relevant information encoded in each (purported) qudit supplied by Bob from the
first lab to the second, and uses the resulting teleported states in
the remainder of the protocol.   Since she prepared the relevant
qudits herself, she can ensure that they carry no information other
than that prescribed by the protocol.

To employ this strategy, Alice clearly must be able to teleport states
at least over short distances.  She may thus simply use the version
of the protocol described above, in which teleportation is used as 
a short secure channel.   If so, of course, she needs only teleport
any given qudit supplied by Bob once (from $P$ to $P'_j$). 
This already guarantees the qudit carries no extra information, 
and so there is no need for a separate initial teleportation in the
neighbourhood of $P$.  

Another possible way of dealing with this issue could be to design device
independent protocols for the tasks discussed here.
Device-independent protocols have been proposed for QKD \cite{BHK,BKP,AGM,
AMP,ABGMPS,PABGMS,McKague,MRC,Masanes,HRW2,MPA,HR}, quantum randomness
expansion \cite{ColbeckThesis,PAMBMMOHLMM,ckrandom} and bit commitment
and coin tossing \cite{DIBC}, among other tasks.  Many of these
protocols have been shown to be vulnerable to attacks
when devices are reused \cite{BCK}, although 
possible defences have also been identified \cite{BCK}.   The 
ultimate scope of device-independent quantum cryptography 
remains an intriguing open question.    
It therefore would be interesting to explore whether device-
independent protocols can be adapted to the relativistic 
cryptographic setting considered here.

\section*{Methods}

\subsection*{Proof details}

We complete here the proof of eq. (\ref{secureN}).

Suppose that Bob makes his measurements on the states in succession
in some order. Denote by $Di_{k}$ the random variable which equals
$1$ (respectively $0$) if Bob's test at site $i$ on returned qudit $k$
succeeds (respectively fails).
Alice passes Bob's test at site $i$ if 
$\sum_{k=1}^{N}Di_{k}\geq \frac{N}{2}\left(1+\frac{2}{d+1} +\epsilon\right)$ 
where the security parameter $\epsilon>0$. 
Let $P_{i}^{N}=\mathrm{Prob}\left(\sum_{k=1}^{N}Di_{k}\geq \frac{N}{2}\left(1+\frac{2}{d+1} +\epsilon\right)\right)$.
To establish security with respect to a dishonest Alice, we need to
bound $P_{1}^{N}+P_{2}^{N}$.

Denote by $Z_{k}=\left(\sum_{j=1}^{k}D1_{j}+D2_{j}\right)\text{\textendash}k\left(1+\frac{2}{d+1}\right)$.
In section \ref{errorsetal}  it was shown that $A$ conditioning on the
measurement results on returned pairs of 
qudits $1\leq i\leq k-1$  cannot increase above $(1 + \frac{2}{d+1})$ 
her success probability of
passing Bob's test on the $k$-th returned pair of qudits. 
From this it follows that
$Z_{k}$ is a supermartingale \cite{gs}. That is $\vert Z_{k}\vert<\infty$
and $\mathrm{E}\left(Z_{k}\vert W_{j}\right)\leq Z_{j}$ for all $k>j$,
where $W_{j}$ is the set of all measurement outcomes up to and including
round $j$. The martingale increments $Z_{k}-Z_{k-1}=D1_{k}+D2_{k} - 
(1+\frac{2}{d+1})$
are bounded in absolute value by $(1+\frac{2}{d+1})$. Hence we can apply the Azuma-Hoeffding
inequality \cite{gs,azuma,hoeffding} to get (for any $\epsilon>0$) 
\[
\mathrm{Prob}\left(\sum_{k=1}^{N}D1_{k}+D2_{k}\geq N\left(1+\frac{2}{d+1}+\epsilon\right)\right)=\mathrm{Prob}(Z_{N}>N\epsilon)<\exp\left[-\frac{N\epsilon^{2}}{2(1+\frac{2}{d+1})^2}\right]
\]

We now use this result to bound $p_{1}^{N}+p_{2}^{N}$. 
We have 
\begin{eqnarray*}
p_{1}^{N}+p_{2}^{N} & = & \mathrm{Prob}\left(\sum_{k=1}^{N}D1_{k}\geq \frac{N}{2}\left(1+\frac{2}{d+1} +\epsilon\right)\right)+\mathrm{Prob}\left(\sum_{k=1}^{N}D2_{k}\geq \frac{N}{2}\left(1+\frac{2}{d+1} +\epsilon\right)\right)\\
 & = & \mathrm{Prob}\left(\sum_{k=1}^{N}D1_{k}\geq \frac{N}{2}\left(1+\frac{2}{d+1} +\epsilon\right)\&\sum_{k=1}^{N}D2_{k}<\frac{N}{2}\left(1+\frac{2}{d+1} +\epsilon\right)\right)\\
 &  & +\mathrm{Prob}\left(\sum_{k=1}^{N}D1_{k}\geq \frac{N}{2}\left(1+\frac{2}{d+1} +\epsilon\right)\&\sum_{k=1}^{N}D2_{k}\geq \frac{N}{2}\left(1+\frac{2}{d+1} +\epsilon\right)\right)\\
 &  & +\mathrm{Prob}\left(\sum_{k=1}^{N}D2_{k}\geq \frac{N}{2}\left(1+\frac{2}{d+1} +\epsilon\right)\&\sum_{k=1}^{N}D1_{k}<\frac{N}{2}\left(1+\frac{2}{d+1} +\epsilon\right)\right)\\
 &  & +\mathrm{Prob}\left(\sum_{k=1}^{N}D2_{k}\geq \frac{N}{2}\left(1+\frac{2}{d+1} +\epsilon\right)\&\sum_{k=1}^{N}D1_{k}\geq \frac{N}{2}\left(1+\frac{2}{d+1} +\epsilon\right)\right)\\
 & \leq & 1+\mathrm{Prob}\left(\sum_{k=1}^{N}D2_{k}\geq \frac{N}{2}\left(1+\frac{2}{d+1} +\epsilon\right)\&\sum_{k=1}^{N}D1_{k}\geq \frac{N}{2}\left(1+\frac{2}{d+1} +\epsilon\right)\right)\\
 & \leq & 1+\mathrm{Prob}\left(\sum_{k=1}^{N}D2_{k}+D1_{k}\geq N\left(1+\frac{2}{d+1} +\epsilon\right)\right)\\
 & \leq & 1+\exp\left[-\frac{N\epsilon^{2}}{2(1+\frac{2}{d+1})^2}\right]\ .
\end{eqnarray*}
QED.

\section*{Discussion}

Each of the strategies discussed here has theoretical and
practical advantages in some plausible potential future scenario. 
The teleportation-based strategies are particularly simple 
and elegant; strategies involving transmitting randomised
states require no pre-distributed entanglement or joint
operations on states; strategies involving physically secure
quantum channels require no operations on quantum states.

We have also discussed schemes involving multiple states 
transmitted over classically extended
quantum channels.  These need only states of small
dimension to be created and manipulated, and need only short range
secure quantum communication.  While even these schemes still pose 
experimental challenges, they seem to us interesting candidates 
for practically implementing the unconditionally secure bit
commitment protocol of Ref. \cite{bcsummoning} and other relativistic
quantum cryptographic tasks over long ranges.  

Note that, in the case of bit commitment, there are also other relativistic
protocols available \cite{kentrelfinite, bcmeasurement}, which rely on 
different principles.  Which of these is most suited to practical 
implementation will depend on the precise application,
as well as on the evolution of technology.  Each has advantages that
may be compelling in some scenarios.  
In particular, the protocol of Ref. \cite{bcsummoning} has the 
advantage that a bit can be committed at a point $P$ and then
persuasively unveiled at a single point $Q$ lightlike separated
from $P$, without waiting for data gathered at other spacelike
separated points in order to verify the unveiling.   
It also has an efficiency advantage in networks where multiple
lightlike transmission directions are possible:  $\log N$ bits can be 
committed by transmitting an unknown qudit in a direction chosen
from $N$ possibilities. 

In summary, we hope the schemes outlined here
will encourage further theoretical and experimental
investigation of the extent to 
which relativistic quantum cryptographic 
tasks can be securely and reliably implemented with
current or forseeable technology.   
From a broader theoretical perspective, our
discussion also illustrates an intriguing application of 
teleportation to cryptography, somewhat different from those
previously considered.   It was realised from the outset
that teleportation has the beautiful cryptographic features that it enables $A$ to
communicate quantum information to $B$ securely, and to do so
even when $A$ does not know $B$'s location. 
We see here another cryptographic feature of teleportation, relying on
the fact that $A$ can securely teleport a state, known to $B$ but
not to $A$, to a location, known to $A$ but not to $B$.     
When light speed delays are negligible, this is not very significant,
since $A$ can teleport between any sites effectively instantaneously.
In relativistic cryptography, though, it means that $A$ can be
constrained -- she must choose one destination from a spacelike
separated set -- and yet conceal information (her choice of
destination) from $B$.   
Our protocols also illustrate that, in quantum relativistic
cryptography, the principle underlying teleportation -- that classical and
quantum communications can advantageously be separated and
recombined at different locations -- has much wider applications.

\acknowledgments
AK thanks Charles Bennett for illuminating discussions of the cryptographic
properties of teleportation and Paul Kwiat for helpful discussions. 
SM and JS thank Stefano Pironio for useful discussions.
AK was partially supported by a Leverhulme Research Fellowship, a grant
from the John Templeton Foundation, and by Perimeter Institute for Theoretical
Physics. Research at Perimeter Institute is supported by the Government of Canada through Industry Canada and
by the Province of Ontario through the Ministry of Research and Innovation.
AK, SM and JS acknowledge the support of the
EU Quantum Computer Science project (contract
255961).  SM and JS acknowledge the support of the Fonds de la Recherche
Scientifique - FNRS and the EU
projects QALGO and DIQIP.

%\bibliography{../bibtex}

\section*{Author Contributions Statement}

The results reported here are collaborative work between AK, SM and
JS.  All authors contributed to writing the manuscript. 
AK prepared Figure 1. All authors reviewed the manuscript and figure. 

\section*{Competing Financial Interests}

The authors declare no competing financial interests.

\section*{Figure Legends}

Figure 1: Illustration, in $1+1$ dimensions (not to scale), of how quantum channels can be securely extended classically.

Alice controls a laboratory
including on its border the points $P$, $P'_1$ and $P'_2$, and
disjoint laboratories including on their borders the points $Q_1$ 
and $Q_2$ respectively, where $P P'_1 Q_1$ and $P P'_2 Q_2$ are
lightlike lines.   Bob may control the rest of space-time.
Alice receives the unknown state $\ket{\psi}$ from Bob at $P$, 
and transmits it securely to $P'_i$ (where she chooses $i =1$ or $2$).
There
she randomizes it and returns the randomized state to Bob.  The
classical data describing how the state was randomized are 
transmitted along a secure classical channel and returned to Bob at $Q_i$.
Alice 
transmits dummy quantum and classical information and returns 
them to Bob at the corresponding points on the opposite
wing

\end{document}